\documentclass[superscriptaddress,aps,twocolumn,secnumarabic,amsmath,amssymb,nofootinbib,nobalancelastpage]{revtex4-1}

\usepackage{graphicx}
\usepackage{url}
\usepackage{natbib}
\usepackage[hypertexnames=false]{hyperref} 	
\usepackage[all]{hypcap}			
\usepackage[english]{babel}
\begin{document}

\title{PyPanda: a Python Package for Gene Regulatory Network Reconstruction}
\author{David G.P. van IJzendoorn}
\affiliation{Department of Pathology, Leiden University Medical Center, Leiden, the Netherlands}
\author{Kimberly Glass}
\affiliation{Channing Division of Network Medicine, Department of Medicine, Brigham and Women's Hospital, Harvard Medical School, Boston, MA, USA}
\author{John Quackenbush}
\affiliation{Department of Biostatistics and Computational Biology, Dana-Farber Cancer Institute, Boston, MA, USA}
\affiliation{Department of Biostatistics, Harvard T.H. Chan School of Public Health, Boston, MA, USA}
\affiliation{Department of Cancer Biology, Dana-Farber Cancer Institute, Boston, MA, USA}
\author{Marieke L. Kuijjer}\email{mkuijjer@jimmy.harvard.edu}
\affiliation{Department of Biostatistics and Computational Biology, Dana-Farber Cancer Institute, Boston, MA, USA}
\affiliation{Department of Biostatistics, Harvard T.H. Chan School of Public Health, Boston, MA, USA}

\begin{abstract}
\textbf{Summary}: PANDA (\underline{P}assing \underline{A}ttributes between 
\underline{N}etworks for \underline{D}ata \underline{A}ssimilation) is a gene 
regulatory network inference method that uses message-passing to integrate 
multiple sources of `omics data. PANDA was originally coded in C++. In this 
application note we describe PyPanda, the Python version of PANDA. PyPanda runs 
considerably faster than the C++ version and includes additional features for 
network analysis. \textbf{Availability and implementation}: The open source PyPanda Python package is freely available 
at \url{http://github.com/davidvi/pypanda}. \textbf{Contact}: \href{d.g.p.van\_ijzendoorn@lumc.nl}{d.g.p.van\_ijzendoorn@lumc.nl}\\
\end{abstract}

\maketitle

\section{Introduction}\label{introduction}
Accurately inferring gene regulatory networks is one of the most important 
challenges in the analysis of gene expression data. Although many methods have 
been 
proposed~\citep{altay2011differential,faith2007large,lemmens2006inferring,
ernst2008semi}, computation time has been a significant limiting factor in their 
widespread use. PANDA (\underline{P}assing \underline{A}ttributes between 
\underline{N}etworks for \underline{D}ata \underline{A}ssimilation) is a gene 
regulatory network inference method that uses message passing between multiple 
`omics data types to infer the network of interactions most consistent with the 
underlying data~\citep{glass2013passing}. PANDA has been applied to understand 
transcriptional programs in a variety of 
systems~\citep{glass2014sexually,glass2015network,lao2015genome}.

Here we introduce PyPanda, a Python implementation of the PANDA algorithm, 
following the approach taken in Glass {\it et al.}~\citep{glass2015high} and optimized for matrix 
operations using NumPy~\citep{van2011numpy}. This approach enables the use of 
fast matrix multiplications using the BLAS and LAPACK functions, thereby 
significantly decreasing run-time for network prediction compared with the 
original implementation of PANDA, which was coded in C++ and used 
for-loops~\citep{glass2015high}. We observe further speed increase over the 
C++-code because PyPanda automatically uses multiple processor-cores through the 
NumPy library. We have also expanded PyPanda to include common downstream 
analyses of PANDA networks, including the calculation of network in- and 
out-degrees and the estimation of single-sample networks using the recently 
developed LIONESS algorithm~\citep{kuijjer2015estimating}.

\section{Approach}\label{approach}
\subsection{Comparing PANDA C++-code to Python-code}\label{comparing}
We compared the C++-code and Python-code versions of PANDA using several 
metrics. First, we assessed the two implementations by comparing the number of 
lines of code. Using the {\it cloc} utility we counted the number of lines of 
C++-code and Python-code. The C++-code counted 1132 lines of code. The 
Python-code counted 258 lines of code, significantly shorter (4.4 times) than 
the C++-code. The Python-code also includes features such as the LIONESS 
equation and in- and out-degree calculation. Without these features 
the Python-code is only 155 lines of code. Because the Python implementation is 
much more concise than the C++-code it is easier to interpret and modify. 

Next we performed a speed comparison test between the C++-code and the 
Python-code. We used built-in timing functions for both languages, directly 
before and after the message passing part of the code as this is the step that 
consumes the most time~\citep{glass2015high}. For the C++-code, we used {\it 
gettimeofday()} to record time in milliseconds before and after the message 
passing algorithm. For the Python code we implemented the {\it time.time()} 
function around the message passing algorithm. The C++-code was compiled using 
the {\it clang} compiler (version 3.8.0) with speed optimization flag -O3. 
Python (version 2.7.10) was used with NumPy (version 1.10.1) using the BLAS and 
LAPACK algebraic functions. All analyses were run on a server running x86\_64 
GNU/Linux.

The speed of the network prediction was tested using simulated networks of 
$Ne=Na$ dimensions, where $Ne$ is the number of effector nodes and $Na$ is the 
number of affected nodes~\citep{glass2015high}. For each of several 
different network sizes ($Ne=Na=125$ to $Ne=Na=2000$ nodes, in steps of 125) we generated 
ten random `motif data' networks according to the method described 
in Glass {\it et al.}~\citep{glass2015high}. We then ran the Python and C++ versions of PANDA 
using these simulated motif data together with identity matrices for the 
protein-protein interaction and co-expression information. For runs 
on the same initial `motif data' networks, we verified that the C++-code and 
Python-code returned exactly the same output network, as expected due to the 
deterministic nature of PANDA.

\begin{figure}[!tbp]
  \includegraphics[width=250px]{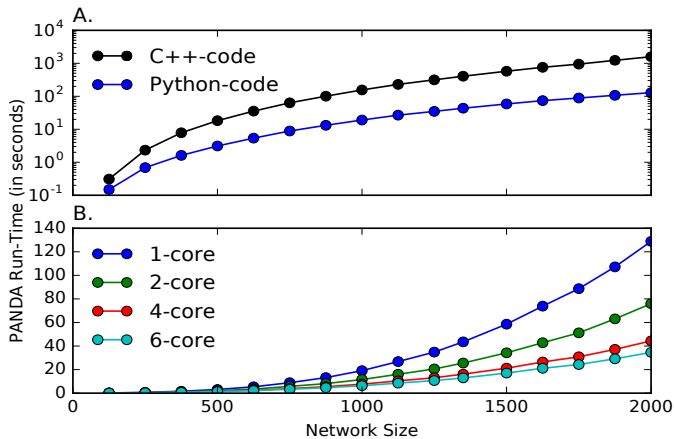}
  \caption{Speed comparison for network reconstruction on networks of different sizes using (A) the C++-code and the Python-code, (B) the Python-code running on a single CPU compared with multicore (6 CPU cores).}
  \label{fig1}
\end{figure}

The C++-code only uses one CPU core. In comparing the C++-code with the 
Python-code using a single core, we found a 2.07-fold speed-up relative to the 
C++-code for the smallest network ($Ne=Na=125$) tested. The speed increase of 
the Python-code over the C++-code became larger as the network size increased. 
For example, the Python-code performed 12.31 times faster for the largest 
network ($Ne=Na=2000$) (Figure~\ref{fig1}A). Recorded run times 
across the ten random networks had a standard  deviation of 0.04s and 2.59s for the smallest ($Ne=Na=125$) and largest ($Ne=Na=2000$) networks, 
respectively using the C++ code. Using the Python code these were reduced to 
0.03s and 0.099s.

Given the abundance of multicore computing resources currently available, we 
also tested the speed increase when running the Python-code on multiple cores 
compared with running the Python-code on a single core. We found that for the 
smallest network the speed was 1.45 times faster when using 6 cores compared with 
using only a single core; for the largest network the speed increase was 
3.7-fold (Figure~\ref{fig1}B). 

This increase in speed enables reconstruction of networks with larger numbers of 
regulators and target genes. For example, using the Python-code significantly 
decreases the time required to infer a human gene regulatory network ($Ne=1000$, 
$Na=20000$), from approximately 18h with the C++-code to only about 2h 
with the Python-code. This speed-up is especially important as transcription 
factor motif databases are frequently updated to include more motifs. Further, 
the decreased running time helps to enable the estimation of network 
significance by making the use of bootstrapping/jackknifing methods much more 
feasible.

\subsection{Additional Features}\label{features}
In addition to reconstructing one regulatory network based on a data set 
consisting of multiple samples, PyPanda can also reconstruct single-sample 
networks using the LIONESS algorithm~\citep{kuijjer2015estimating}. In PyPanda, 
the LIONESS method uses PANDA to infer an `aggregate' network representing a 
set of $N$ input samples, infers a network for $N-1$ samples, and then applies a 
linear equation to estimate the network for the sample that had been removed. 
The process is then repeated for each sample in the original set, producing $N$ 
single-sample networks. PyPanda can also use LIONESS to reconstruct 
single-sample networks based on Pearson correlation.

PyPanda also includes functions to calculate in-degrees (the sum of edge weights 
targeting a specific gene) and out-degrees (the sum of edge weights pointing out 
from a regulator to its target genes). These summary metrics can be used for 
downstream network analysis~\citep{glass2014sexually}.
%

\section{Conclusion}\label{conclusion}
PANDA is a proven method for gene regulatory network inference but, like most 
sophisticated network inference methods, its runtime has limited its utility. 
The Python implementation of PANDA uses matrix operations and incorporates the 
NumPy libraries, resulting in a significant simplification of the code and a 
dramatic increase in computing speed, even on a single processor. When applied 
to a test data set and run on multiple processing cores, this increase in speed 
was even greater, decreasing processing times by a factor of 46 relative to 
the original C++-code. This creates opportunities to greatly expand the use of 
PANDA and to implement additional measures of network significance based on 
bootstrapping/jackknifing. PyPanda also includes the LIONESS method, which 
allows inference of single-sample networks, as well as a number of other useful 
network metric measures. The open source PyPanda package is freely available at 
\url{http://github.com/davidvi/pypanda}.

\section*{Acknowledgments}
The authors would like to thank Judith V.M.G. Bov\'ee, MD, PhD and Karoly 
Szuhai, MD, PhD for thoughtful discussions and Cho-Yi Chen, PhD for testing PyPanda.

\section*{Funding}
This work has been supported by the National Institutes of Health [R01 HL111759 
to J.Q., K.G., P01 HL105339 to J.Q., K.G., M.L.K.] and Leiden University Fund 
[5259/4-6-2015/Gg to D.G.P.IJ].
%
\bibliographystyle{apsrev-title}
\bibliography{pypandabib_v2} 

\end{document}